\documentclass{optica-article}

\journal{opticajournal} 

\articletype{Research Article}

\usepackage{lineno}

\begin{document}

\title{Single picture single photon single pixel 3D imaging through unknown thick scattering medium}

\author{Long Pan,\authormark{1,2,$\dag$} Yunan Wang,\authormark{1,$\dag$} Yijie Lou,\authormark{1,$\dag$} and Xiaohua Feng\authormark{1,3} }

\address{\authormark{1}Research Center for Frontier Fundamental Studies, Zhejiang Laboratory, Hangzhou, 311121, China\\}

\email{\authormark{2}panlong@zhejianglab.com\\
\authormark{3}xiaohuafeng@zhejianglab.com\\ 
\authormark{\dag}These authors contributed equally to this work}

\begin{abstract*} 
Imaging through thick scattering media presents significant challenges, particularly for three-dimensional (3D) applications. This manuscript demonstrates a novel scheme for single-image-enabled 3D imaging through such media, treating the scattering medium as a lens. This approach captures a comprehensive image containing information about objects hidden at various depths. By leveraging depth from focus and the reduced thickness of the scattering medium for single-pixel imaging, the proposed method ensures robust 3D imaging capabilities. We develop both traditional metric-based and deep learning-based methods to extract depth information for each pixel, allowing us to explore the locations of both positive and negative objects, whether shallow or deep. Remarkably, this scheme enables the simultaneous 3D reconstruction of targets concealed within the scattering medium. Specifically, we successfully reconstructed targets buried at depths of 5 mm and 30 mm within a total medium thickness of 60 mm. Additionally, we can effectively distinguish targets at three different depths. Notably, this scheme requires no prior knowledge of the scattering medium, no invasive procedures, reference measurements, or calibration. 

\end{abstract*}

\section{Introduction}
Scattering and absorption are the two great obstacles induced by scattering medium when light propagates through. They chaos the plenoptic function of the light, which makes imaging into/through scattering medium tough. However, imaging into/through scattering medium is inevitable and indispensable in some field like astronomy \cite{fugate1991measurement}, remote sensing \cite{chen2013ghost}, medical imaging \cite{zhao2021high}, autonomous driving \cite{liu2022multi}. Considerable literature paid attention to addressing this issue and 2D imaging is first focused \cite{bertolotti2012non,bertolotti2022imaging,sdobnov2024advances,jian2021underwater,ogunrinde2021review}. More than a decade ago, related works mainly focused on optically thin scattering medium (less than 5 scattering mean free paths) or weak scattering \cite{narasimhan2005structured,ntziachristos2010going}. 3D imaging capability was paid little attention and underdeveloped. In the past decade, imaging into/through thick scattering medium and 3D imaging has been becoming popular with the development of research methods and practical requirements \cite{kang2015imaging,lyons2019computational,lee20223d,shi2022non}. Whereas some works pay attention to these field, neither the reconstruction quality is that satisfying \cite{lyons2019computational} nor the penetrating depth is that deep \cite{shi2022non}. In addition, calibration, prior knowledge of the scattering medium and objects, invasion and huge amount of calculation are factors assailing imaging through/into scattering medium.

Typical methods for imaging into/through scattering medium can be classified into following kinds: speckle correlation imaging \cite{bertolotti2012non,katz2014non}, wave-front shaping based imaging \cite{vellekoop2007focusing, mosk2012controlling}, transmission matrix based imaging \cite{popoff2010measuring,popoff2010image}, point spread function (PSF) based imaging \cite{metari2007new,antipa2017diffusercam}, transfer/diffusion equation based imaging \cite{lyons2019computational,du2022boundary}, single pixel imaging \cite{tajahuerce2014image}, phase-space imaging \cite{takasaki2014phase,liu20153d}, and other traditional methods. Recently, with the rapid increase of computing power, data-driven deep learning \cite{li2018imaging} and physics-data-driven deep learning \cite{zhu2021imaging,zhang2023physical,xue2024fully} show their strength in scattering scenarios.  

Speckle correlation imaging, one of the most popular methods for imaging through/into scattering medium, utilizes optical memory effect (angular memory effect) of the scattering medium. Traditional memory effect based methods for imaging through/into scattering medium, such as \cite{bertolotti2012non,katz2014non}, can only obtain 2D reconstruction results. Besides, memory effect, which limits the field of view and penetrating depth, likes a double-edged sword. Combining speckle correlation and other methods, 3D reconstruction can be obtained although generally with limited field of view and shallow penetrating depth. A temporal method using coherence gating enables speckle correlation 3D imaging \cite{salhov2018depth}. Parallax is another approach for speckle correlated 3D imaging \cite{shi2017non, wang2024imaging}. PSF manipulation boosts the  speckle correlated 3D imaging as well \cite{singh2017exploiting,xie2018extended}. Recently, 3D memory effect (angular memory effect and axial memory effect) is found and facilitates 3D imaging through scattering medium \cite{shi2022non,aarav2024depth}.

Phase-space imaging exploits 4D phase-space measurements to mitigate the effects of scattering and obtains 3D image with redundant \cite{takasaki2014phase,liu20153d,liu2017multiplexed}. Structured illumination has long been used for underwater 3D imaging although the penetrating depth is short and the scattering is not strong \cite{narasimhan2005structured}. Combined with computed tomography, structured illumination shows another approach to boost 3D imaging through/in scattering medium \cite{kristensson2012quantitative}. With time-of-flight method, several schemes developed from diffuse tomography, are proposed for 3D imaging into scattering medium \cite{lindell2020three, du2022boundary,deng2023scan,du2023non}. Structured illumination and spatial filtering is utilized to assist 3D imaging through scattering medium \cite{kristensson2012quantitative,jauregui2019single,zhao2022imaging}.

Besides above mentioned methods based on 2D array detectors, single pixel imaging, which requires only one detector, has been found the advantage of reduced effective scattering length \cite{zhang2015single}. The scattering medium in the illumination light path greatly impacts the reconstruction quality. However, the scattering medium between imaging objects and detecting sensor has almost no effect on the deterioration of image quality expect weakening light intensity. Time gating which decreases the multi-scattering photons by gating the photon travels not within the gate width is one of several-scattering photon methods. It has been demonstrated a boosted resolution of imaging through scattering medium \cite{selb2006time}. 

Depth from focus (DFF) is a depth estimation method applicable to free space \cite{xiong1993depth,subbarao1998selecting}. By calculating the pixel-level contrast from a serious of 2D pictures via metric, the depth is obtained. The geometry of the imaging optics should be known. Depth from focus requires the acquired pictures with high quality.

Here, we introduce a novel 3D imaging model, termed single picture single photon single pixel 3D imaging (SP$^3$ 3DI), for imaging through unknown thick scattering medium. SP$^3$ 3DI incorporates single photon detection, time gating, single pixel imaging, and depth from focus. Single photon detection feature enhances the sensitivity of SP$^3$ 3DI. Time gating effectively filters out multi-scattering photons, thereby improving resolution. Single pixel imaging enables a reduced scattering length. Depth from focus offers the depth information. Different from traditional DFF methods, SP$^3$ 3DI requires only one picture for DFF. By treating scattering medium as a lens focusing the entrance plane of the scattering medium, SP$^3$ 3DI takes a blurred image that contains object information from multiple depths. Our algorithms then reconstruct the focal stack and depth information. We have developed both traditional and deep learning-based algorithms for this purpose. Furthermore, we demonstrate the capability of 3D imaging through scattering medium at both shallow and deep simultaneously. It should be noted that, although time gating is used, the depth information is not related to time of flight.

\section{System Overview}
\begin{figure*}[p]
\centering
\includegraphics[scale=0.45]{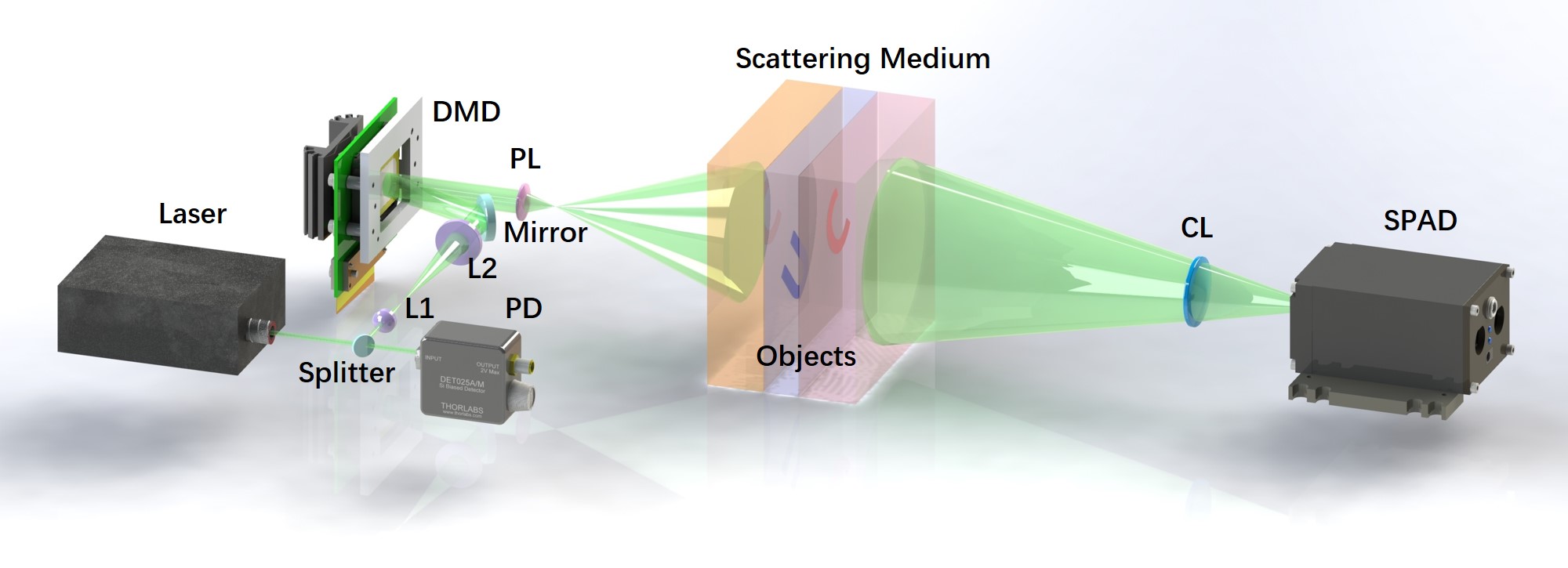}
\caption{Schematic of SP$^3$ 3DI. PD: photodiode, Mr: mirror, DMD: Digital Micromirror Device, PL: Projecting Lens, CL: collective lens, SPAD: Single Photon Avalanche Diode.}
\label{fig:setup}
\vspace{-10pt}
\end{figure*}

The schematic of proposed SP$^3$ 3DI is shown in Fig. \ref{fig:setup}. A 532 nm wave length picosecend laser source ($\sim$10 ps pulse width, 1.0 W average power, 1 MHz repetition rate) emits light beam. After the splitter, 10$\%$ light entrance a high speed photodiode (PD, \emph{Thorlabs, DET025A/M})  working as the synchronization for the photon measurement. 90$\%$ light is expanded by lens L1 and L2, and is reflected by a mirror before illuminating the digital micromirror device (DMD, \emph{Texas Instruments, V-7001, Vialux Gmbh}). DMD spatially modulates the light beam in each exposure with sequenced patterns ($A^m$, where $A$ stands for the patterns, $m$ denotes the $m$th pattern). Projecting Lens (PL) images the modulated light beam into scattering medium. The objects are hidden in the scattering medium. The transmitted photons are collected by a collective lens (CL) and sent to a a single photon avalanche diode (SPAD, \emph{Micro Photon Device, FastGated SPAD}). SPAD detects the low flux photons and is triggered one time at most during one laser pulse period. A time-correlated single photon counter (TCSPC, \emph{PicoQuant, Timeharp260, single}) records the signal $y^m$ from SPAD with the synchronization.  

As the modulated light beam penetrates the scattering medium, the patterns gets blurred. Previous works and our work have demonstrated that the convolution kernel can be approximated by Gaussian kernels with different scales in different depths. The objects locates in depth $z_i$ is illuminated by the blurred patterns with kernel size $G_i$. Considering the randomness of the photon statistics, the accumulated temporal profile of the photons signal at the SPAD after multi laser pulse can be approximately described to be
\begin{eqnarray}
{y^m}(t) = P(t - mT)\sum\limits_i {\int {A_{bi}^m{x_i}d{z_i}} } + \varepsilon^m(t)   ,
\label{eq:temporal y signal}
\end{eqnarray}
where $A_{bi}^m$  and ${x_i}$ denotes the blurred pattern and object hidden at depth  $z_i$, respectively. $A_{bi}^m = {A^m} * {G_i}$, and $*$ stands for convolution. Here, noise term $\varepsilon^m(t)$ is considered. It should be noted that the blurring kernel is of the form ${G_i} = \exp \{ - \frac{{{r}^2}}{{2\sigma_i }}\} $ and ${G_i} = {G_{ - i}}$ and $\sigma_i$ is the depth related blurring kernel scale. $T$ stands for period of the laser pulse. $P(t - mT)$ denotes the temporal profile of the laser pulse transmitting the scattering medium. The temporal waveform of the received signal ${y^m}(t)$ is a broadened form of the emitted laser pulse. 

Time gating is proven to reduce the multi-scattering components in the receiving photon signal. The later arrived part in each ${y^m}(t)$ can be recognised as the  multi-scattering components and is gated out. And the bullet photons and the several-scattering photons are kept as effective signal photons. The time gated measurement signal for single pixel imaging is
\begin{eqnarray}
{y^m} = \int_{(m - 1)T}^{(m - 1)T + \Delta t} {P(t - mT)dt} \sum\limits_i {\int {({A^m} * G_i){x_i}d{z_i}} } + \varepsilon _g^m ,
\label{eq:y signal gated}
\end{eqnarray}
where $\Delta t$ is the gate width. ${\varepsilon _g}$ is the time gated noise term and $\varepsilon _g^m = \int_{(m - 1)T}^{(m - 1)T + \Delta t} {{\varepsilon ^m}(t)dt}  $. After some calculation, Eq. \ref{eq:y signal gated} can be transformed to
\begin{eqnarray}
{y^m} = \int_{(m - 1)T}^{(m - 1)T + \Delta t} {P(t - mT)dt} \int {{A^m}\sum\limits_i {({G_i} * {x_i})} d{z_0}} + \varepsilon _g^m , 
\label{eq:y signal }
\end{eqnarray}
where $z_0$ is the intersection between the scattering medium and the input modulated pattern. ${G_i} * {x_i}$ is the blurred image of the object $x_i$ back-propagated to the intersection plane.

\section{Methods}
\begin{figure*}[p]
	\centering
	\includegraphics[width=1\textwidth]{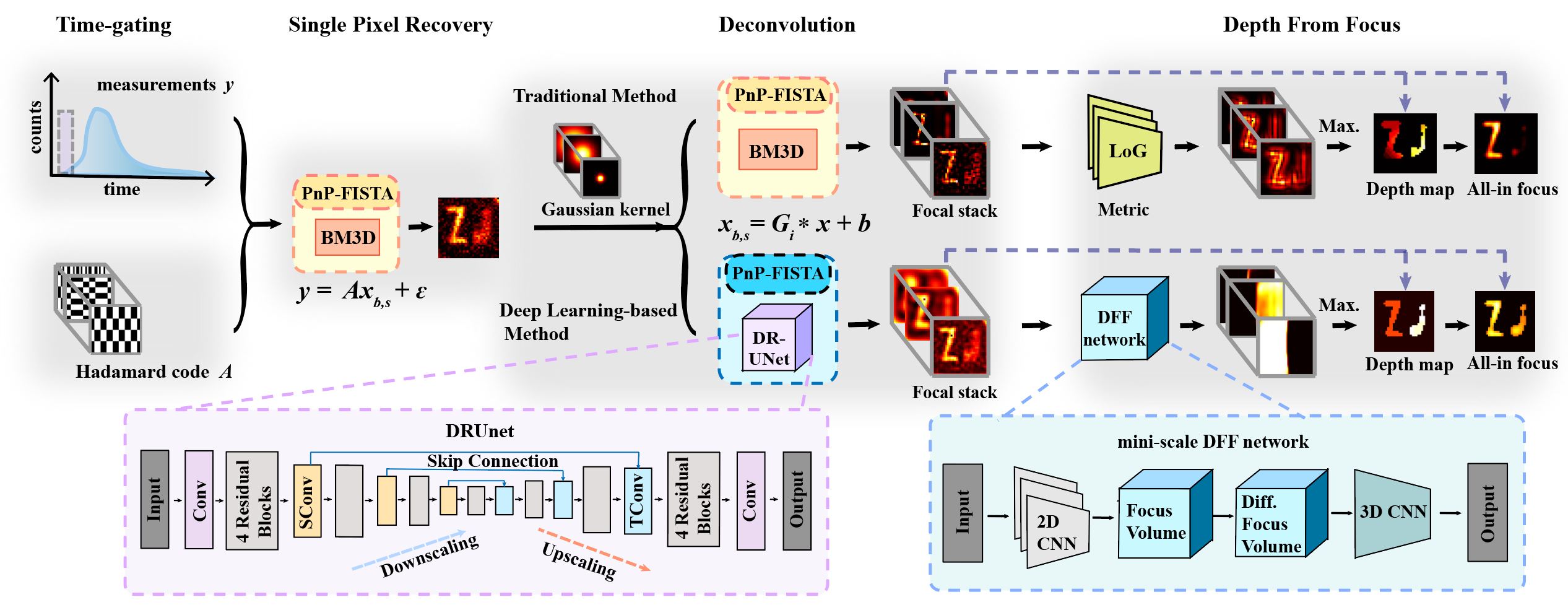} %
	\caption{Workflow of SP$^3$ 3DI. Measurements $y$ is first time gated. Pulg-and-play fast iterative shrinkage-thresholding algorithm (PnP-FISTA) is then combined with block-matching and 3D filtering (BM3D) to retrieve the blurred image of objects. Traditional method is based on PnP-FISTA and BM3D to deconvolve the blurred image before obtaining focal stack. Laplacian of Gaussian (LOG) works as the metric to evaluate the depth of each pixel. Deep learning-based method utilizes PnP-FISTA and deep ResNet and U-Net (DRUNet) to reconstruct the focal stack. Depth from focus (DFF) network evaluates the depth of each pixel.} 
	\label{fig:algorithms}
 \vspace{-10pt}
\end{figure*}

Considering the fact that the PSF of the incident light is Gaussian-like and it holds laterally shift-invariant for approximately homogenous scattering medium \cite{lyons2019computational}, the image retrieved from single pixel imaging is an image blurred by Gaussian kernels of different scales in consistent with the object depths. It should be noted that in the regularized optimization problem, images are in vectorized form. The scattering medium can be treated as a lens focusing the entrance plane of the illumination light. Each object locating different depth is blurred by the according blurring kernel. Thus the image in the focal plane of the collective lens can be recognized as a summation of images from different depths. To obtain the 3D image of the objects, two algorithms, based on tradition framework of DFF and deep learning, are proposed. The workflow of two algorithms is shown in  Fig. \ref{fig:algorithms}. The main process of the algorithms can be classified into 4 parts: time gating, single pixel recovery, deconvolution, and depth from focus. The time gated measurements $y$ and code $A$ are fed into single pixel recovery process to reconstruct the blurred image of objects in the entrance plane of scattering medium. By solving the following regularized optimization problem, the summation of the blurred images ${x_{b,s}} = \sum\limits_i {({G_i} * {x_i})} $ is obtained:
\begin{eqnarray}
\mathop {\arg \min }\limits_{{x_{b,s}}} {\left\| {y - A{x_{b,s}}} \right\|^2} + \alpha {\left\| {{x_{b,s}}} \right\|_{TV}}     , \label{eq: regularized problem of ${x_{b,s}}$ }
\end{eqnarray}
where $\alpha $ is the regularization parameter, ${\left\|  \bullet  \right\|^2}$ and ${\left\|  \bullet  \right\|_{TV}}$ denotes the Frobenius and total variation norm, respectively. Pulg-and-play fast iterative shrinkage-thresholding algorithm (PnP-FISTA) \cite{ahmad2020plug} is exploited as the main deconvolution algorithms. As a popular image denoising method, block-matching and 3D filtering (BM3D) \cite{dabov2007image} works as denoiser for single pixel recovery. After single pixel recovery process, deconvolution and depth from focus process are different for two algorithms. Detailed process for two algorithms will be descried.

\subsection{Traditional method}
To further get the deblurred 3D image from ${x_{b,s}}$, parametric Gaussian approximation is used to identify the correct scale of the Gaussian kernel. For object $x_i$, the correct scale of the Gaussian kernel is $G_i$. Since $x_{b,s}$ is a summation of all the objects located in different depths, the image $x_i$ will appear when a series of Gaussian kernel scales is utilized to solve ${x_{b,s}}$. The probelm becomes how to discriminate the correct image of $x_i$ among a stack of images which is corresponding to different depth. The images of different depth form a cubic in which the 2D slices show the image evolution of each depth from blurry to clear. Inspired by the autofocus process of cameras, image metric is utilized for guiding the research of the scales. First, the local absolute contrast for each pixel at different depth is evaluated. Then, the maximum of the lateral pixel corresponds to the correct scale of the Gaussian kernel. To solve $x_{b,s}$ for image $x_i$ in different depth, the deconvolution step by regularization is
\begin{eqnarray}
\mathop {\arg \min }\limits_{{x_i}} {\left\| {{x_{b,s}} - {G_i} *{x_i}} \right\|^2} + \beta {\left\| {{x_i}} \right\|_{TV}}  , 
\label{eq: regularized problem of $x_i$ }
\end{eqnarray}
where $\beta $ is the regularization parameter. And the metric for each lateral pixel $(u,v)$ and depth $z_i$ is
\begin{eqnarray}
{Q_i^{u,v}} = Max\left\{ {\left| {Lo{G_h}(x_i^{u,v})} \right| + \left| {Lo{G_v}(x_i^{u,v})} \right|} \right\}  , 
\label{eq: metric }
\end{eqnarray}
where $LoG_h$ and $LoG_v$ are Laplacian of Gaussian operator of the horizontal and vertical dimension respectively. The scale of the maximum of $Q$ for each pixel is the corresponding depth $z_i$.
\subsection{Deep learning-based method}
Deep ResNet and U-Net (DRUNet) is a powerful deep denoiser which is typically designed for Plug-and-Play image restoration \cite{zhang2021plug}. Its architecture combines the ResNet and U-Net which enable the network to effectively handle complex details in image restoration. In our deep learning-based framework, PnP-FISTA and DRUNet form a deconvolution process. A serious of Gaussian kernel with different scales is used. After deconvolution, a focal stack is obtained. To get the right depth, we explore deep-learning-based depth-from-focus (deep DFF) methods \cite{hazirbas2019deep,yang2022deep}, which recently have been proven to be an effective way to identify the best-focused images. Traditional DFF methods experience significant performance degradation when the image is textureless or signal-to-noise ratio is low. Different from the traditional DFF, the deep DFF model learns from the data distribution. Thus context information is utilized to better identify the best-reconstructed image. 

We use the multi-scale DFF architecture proposed in the Ref. \cite{yang2022deep}, where the differential focus volume module plays a crucial role by integrating both depth and contextual information for depth estimation. Given the focal stack containing reconstructed images, the model predicts the probability that each image is the best -reconstructed. Although the pretrained model can not be directly used in our work due to the huge data discrepancies, we address this by preparing the training data according to our experimental setup, which is generated from the handwritten MNIST dataset. First, We convolve multiple handwritten images with Gaussian kernels of varying scales to mimic the imaging process of objects at different depth within a scattering medium. Next, we concatenate the convolved images and introduce white noise to achieve an appropriate signal to noise ratio (SNR). Finally, we use the PnP framework to perform deconvolution with a series of Gaussian kernels to obtain the focal stack. 
\section{Experimental Results}
To verify the feasibility of the proposed method, the scattering medium is simulated by 3D printed slabs with different thickness to investigate the performance. Targets with positive ones (transmission mask) and negative ones (black tape) are sandwiched between slabs. The thickness of the last slab is 30 mm to guarantee the scattering medium is thick. The height and weight of the slabs are 200mm. Two (positive or negative) targets buried in the same depth and different depth are both experimentally examined. The illumination area on the first slab is about 120 mm $\times$ 120 mm. The patterns are drawn from Hadamard code with a total number of the patterns 1024 which makes the size of the image dimension 32 $\times$ 32. Each entry of Hadamard code is madu up of 24 $\times$ 24 DMD pixels. Considering the DMD projection, the angle between the illuminating patterns and slab is about 45${\rm{^\circ }}$. The time bin of the TCSPC is set to 50 ps. Since the change of the whole thickness of the scattering medium, the pulse width and delay differs in consistence with the thickness of the scattering medium. Because of time gating technique, the uprising part of the photon histogram by the TCSPC is chosen as effective signal. Given the total thickness of the scattering material ranging from 35 mm to 60 mm, the exposure of each pattern changes from 100 ms to 400 ms accordingly. And the laser output power is tuned accordingly to ensure the maximum of the measurements around several hundred. As a consequence, the SNR of measurements will not differ so much that the imaging performance will get affected severely. Besides, the kernel scale is set to range from 0.1 to 4 with a step size 0.1. Thus the reconstructed depth is in the range from 1 to 40 with a step size 1. To denote scattering medium, value 0 in the reconstructed intensity map and depth map is chosen.

\subsection{Single depth}
\begin{figure}[p]
\centering
\includegraphics[scale=0.6]{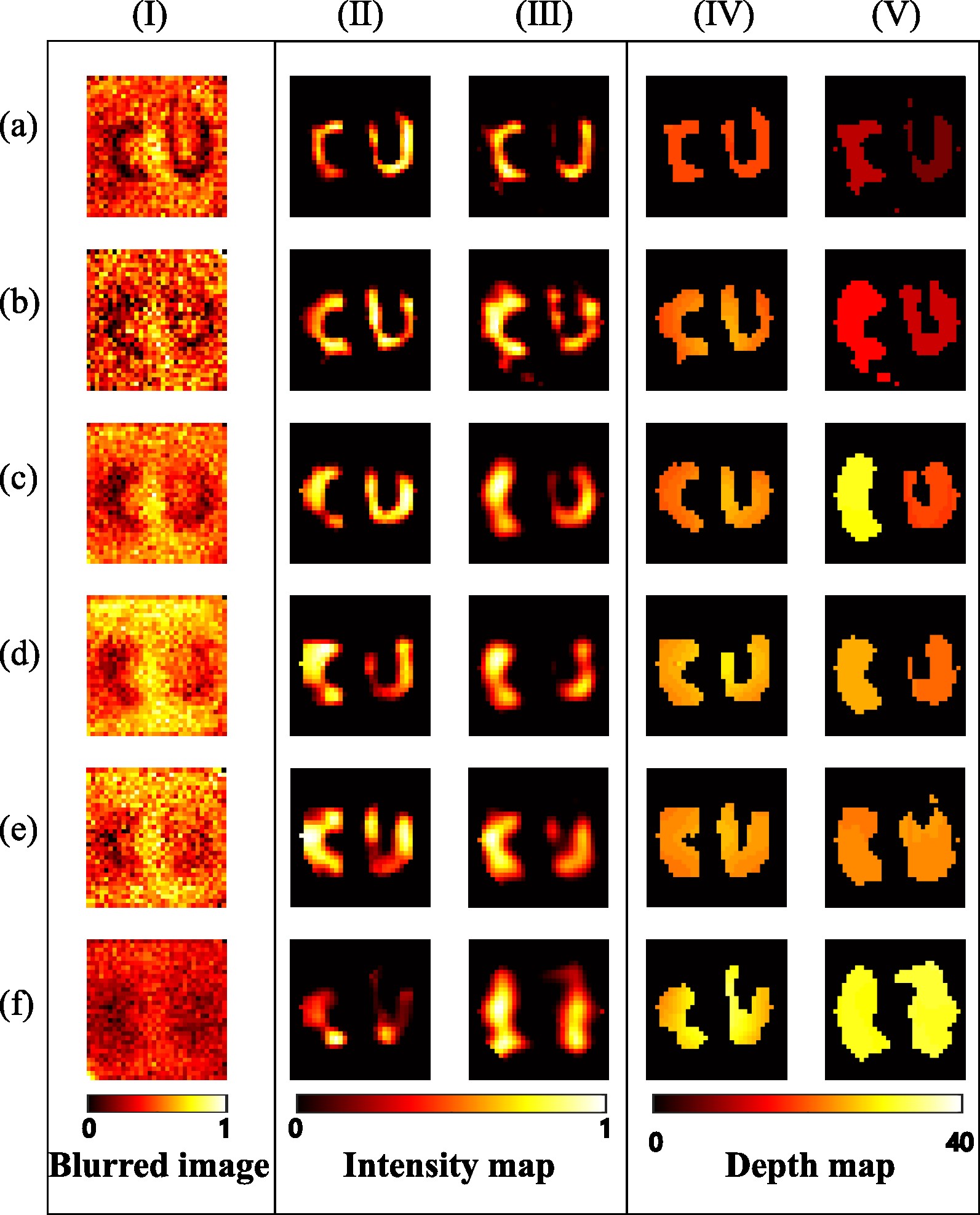}
\caption{Experimental results with negative targets hidden in the same depth. From top to bottom, row (a) to row (f) shows the reconstructed images of the objects hidden in the depth of 5 mm to 30 mm in every 5 mm. Column (I) shows the single pixel recovery result. Column (II) and (III) shows the intensity depth retrieved from traditional methods and deep learning-based method, respectively. Column (IV) and (V) shows the intensity depth retrieved from traditional methods and deep learning-based method, respectively. Value 0 in intensity map and depth map denotes scattering medium.}
\label{fig:negative_same_depth}
\vspace{-10pt}
\end{figure}

In order to demonstrate the depth discrimination of the proposed method, imaging with negative objects is first explored. The letters C and U are attached to a slab with 5 mm thickness and sandwiched by a slab with 30 mm thickness. When changing the effective scattering thickness (buried depth of letters C and U), extra slab with proper thickness is planted. The results are illustrated in Fig. \ref{fig:negative_same_depth}. Two letters C and U are buried in the same depth among 3D printed slabs. From top to bottom, Fig. \ref{fig:negative_same_depth} row (a) to (f) shows the reconstructed maps of letters C and U buried in depth from 5 mm to 30 mm in every 5 mm. Fig. \ref{fig:negative_same_depth} column (I) shows the blurred image reconstructed from different depths. Fig. \ref{fig:negative_same_depth} column (II) and (III) gives the intensity map retrieved from traditional method and deep leaning based method, respectively. Fig. \ref{fig:negative_same_depth} column (IV) and (V) illustrates the depth map reconstructed from traditional method and deep leaning based method, respectively. The quality of the blurred image reconstructed from single pixel imaging become worse as the penetrating depth increases in Fig. \ref{fig:negative_same_depth} column (I) from top to bottom. Even hardly anything can be recognized as the blurred kernel size grows to some extent in Fig. \ref{fig:negative_same_depth} column (I). For both traditional method and deep learning-based method, the performance of intensity map and depth get worse as the effective penetrating depth becomes deeper. The reconstructed line width of letters C and U thickens and the identifiability deteriorates according to Fig. \ref{fig:negative_same_depth}. Comparing column (II)  and (III) in Fig. \ref{fig:negative_same_depth}, traditional method shows better performance of intensity map. While, deep learning base method illustrate better depth consistency in Fig. \ref{fig:negative_same_depth} column (IV) and (V). In addition, traditional method shows better penetrating depth. It should be noted that because the number of gray scales for intensity map is 256 for $0\sim1$ while that for depth map is 41 for $0\sim40$, the depth map may show detailed profile of the objects. 

\begin{figure}[p]
\centering
\includegraphics[scale=0.6]{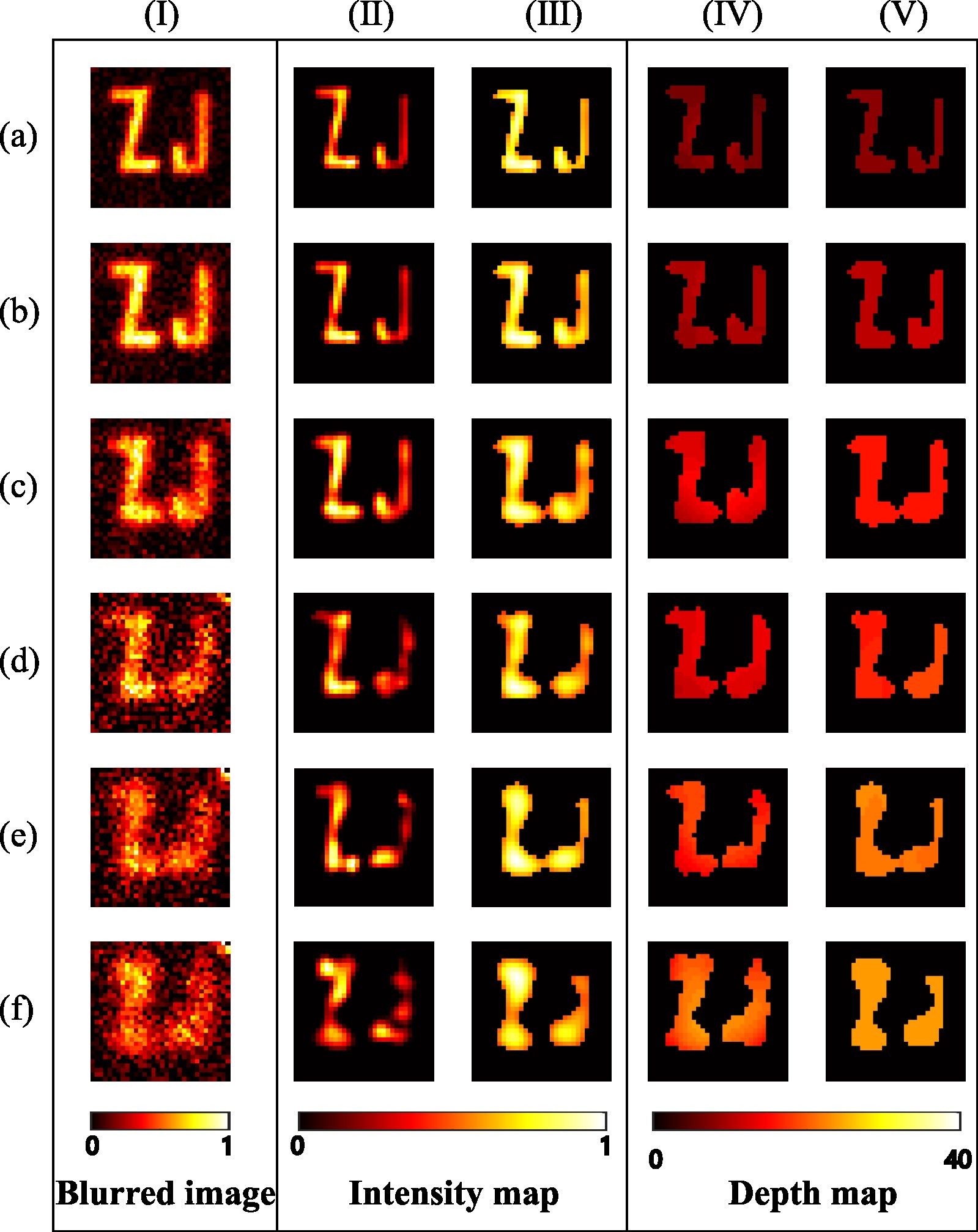}
\caption{Experimental results with positive targets hidden in the same depth. From top to bottom, row (a) to row (f) shows the reconstructed images of the objects hidden in the depth of 5 mm to 30 mm in every 5 mm. Column (I) shows the single pixel recovery result. Column (II) and (III) shows the intensity depth retrieved from traditional methods and deep learning-based method, respectively. Column (IV) and (V) shows the intensity depth retrieved from traditional methods and deep learning-based method, respectively. Value 0 in intensity map and depth map denotes scattering medium.}
\label{fig:positive_same_depth}
\vspace{-10pt}
\end{figure}

3D Imaging with positive objects buried in the same depth is shown in Fig. \ref{fig:positive_same_depth}. Similar to negative targets, letters Z and J are hidden in the same depth. The depth changes from 5 mm to 30 mm in every 5 mm. From top to bottom, Fig. \ref{fig:positive_same_depth} row (a) to (f) shows the reconstructed maps of letters Z and J placed in depth from 5 mm to 30 mm in every 5 mm. Fig. \ref{fig:positive_same_depth} column (I) gives the blurred images recovered from single pixel imaging. From top to bottom, the image quality goes bad in Fig. \ref{fig:positive_same_depth} column (I). Column (II) and (III) in Fig. \ref{fig:positive_same_depth} shows the reconstructed intensity map for traditional method and deep learning-based method, respectively. The imaging quality becomes worse as the depth gets thicker. However, both methods performs better than that in Fig. \ref{fig:positive_same_depth}. Column (IV) and (V) in Fig. \ref{fig:positive_same_depth} reveals the reconstructed depth map for traditional method and deep learning-based method, respectively. It shows a satisfying results for both methods. But deep learning-based method indicates better depth homogeneity for each letter Z and J. By and large, compared with results of negative objects, the intensity map and depth map of positive targets show better performance.

\subsection{Two depths}
\begin{figure}[p]
\centering
\includegraphics[scale=0.6]{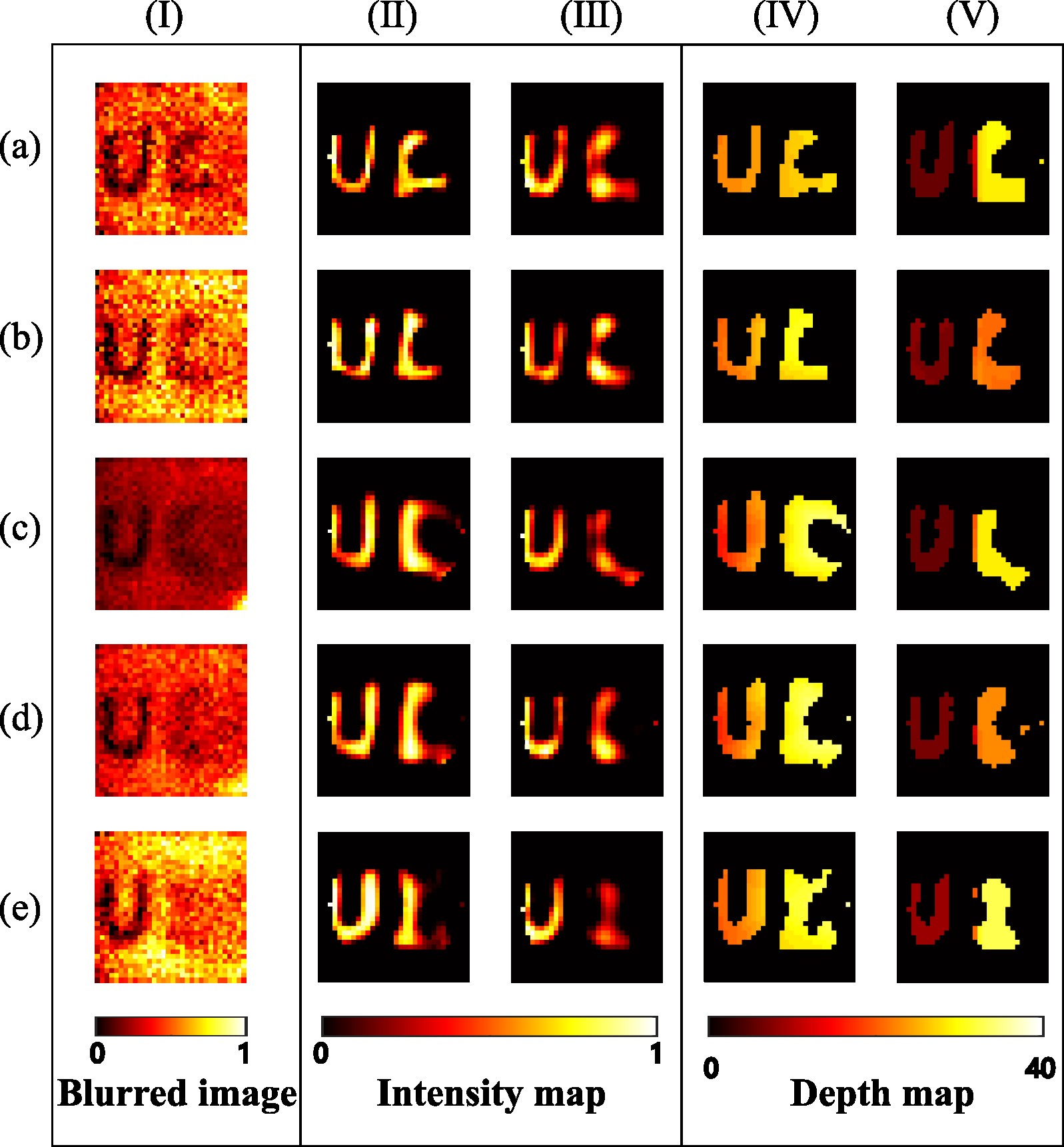}
\caption{Experimental results with negative targets hidden in different depths. From top to bottom, row (a) to row (e) shows the reconstructed images for letter Z fixed in 5 mm depth and letter J hidden in the depth of 10 mm to 30 mm in every 5 mm. Column (I) shows the single pixel recovery result. Column (II) and (III) shows the intensity depth retrieved from traditional methods and deep learning-based method, respectively. Column (IV) and (V) shows the intensity depth retrieved from traditional methods and deep learning-based method, respectively. Value 0 in intensity map and depth map denotes scattering medium.}
\label{fig:negative_different_depth}
\vspace{-10pt}
\end{figure}

3D imaging results of negative targets buried in different depths are presented in Fig. \ref{fig:negative_different_depth}. The scenario consists of two letters U and C. Letter U is buried in depth of 5 mm. The depth of the letter C ranges from 5 mm to 30 mm with a interval 5 mm. Fig. \ref{fig:negative_different_depth} row (a) to (e) shows the reconstructed images as depth increases from top to bottom. The intensity maps and depth maps for both traditional method and deep learning-based method is shown in Fig. \ref{fig:negative_different_depth} column (II) to (V). Column (II) and (IV) in Fig. \ref{fig:negative_different_depth} shows the intensity map results for traditional method and deep learning method, respectively. Column (IV) and (V) in Fig. \ref{fig:negative_different_depth} shows the depth map results for traditional method and deep learning method. Because photons from letter C experience more effective scattering, letter C shows a relative worse reconstruction quality compared with letter U. As letter C buried deeper, letter U strongly suppress letter C from the point of SNR. As the depth difference between letter U and C increases, the depth difference becomes distinct according to Fig. \ref{fig:negative_different_depth} Column (IV) and (V). 

\begin{figure}[p]
\centering
\includegraphics[scale=0.6]{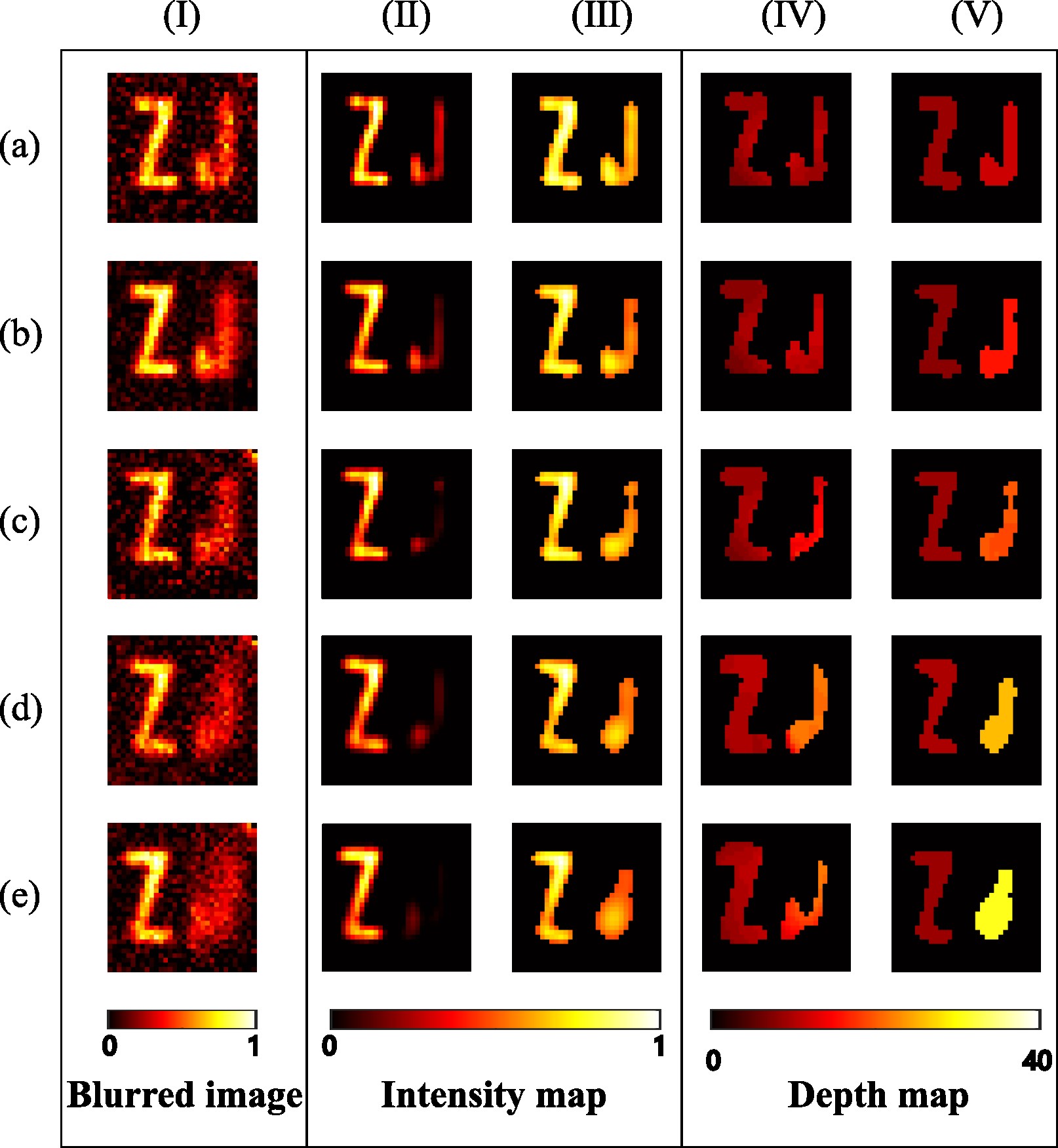}
\caption{Experimental results with positive targets hidden in different depths. From top to bottom, row (a) to row (e) shows the reconstructed images for letter Z fixed in 5 mm depth and letter J hidden in the depth of 10 mm to 30 mm in every 5 mm. Column (I) shows the single pixel recovery result. Column (II) and (III) shows the intensity depth retrieved from traditional methods and deep learning-based method, respectively. Column (IV) and (V) shows the intensity depth retrieved from traditional methods and deep learning-based method, respectively. Value 0 in intensity map and depth map denotes scattering medium.}
\label{fig:positive_different_depth}
\vspace{-10pt}
\end{figure}

The imaging results of positive target hidden in different depth are presented in Fig. \ref{fig:positive_different_depth}. Similar to the case of negative targets, positive target letter Z is buried in a fixed depth 5mm. While positive target letter J is hidden in the depth 10 mm to 30 mm with an interval 5 mm. Fig. \ref{fig:positive_different_depth} row (a) to (e) illustrates the according reconstruction result. Fig. \ref{fig:positive_different_depth} column (I) reveals that the letter Z is less blurred and can be overall reconstructed from single pixel recovery. While letter J turns to be unrecognizable with the increase of hidden depth. Fig. \ref{fig:positive_different_depth} column (II) and column (III), column (IV) and column (V) gives the intensity map and depth map for traditional method and deep learning-based method, respectively. The component of letter J is distinctly suppressed by that of letter Z for traditional method in Fig. \ref{fig:positive_different_depth} column (II). Because the letter J is buried deeper than letter Z, thus, the effective energy from letter Z is stronger than letter J. As for the deep learning-based method in Fig. \ref{fig:positive_different_depth} column (III), the intensity suppression is not that distinct. Fig. \ref{fig:positive_different_depth} column (IV) and (V) show good depth discrimination for both methods. While traditional method show slightly better image detail.

\subsection{Three depths}
\begin{figure}[p]
\centering
\includegraphics[scale=0.6]{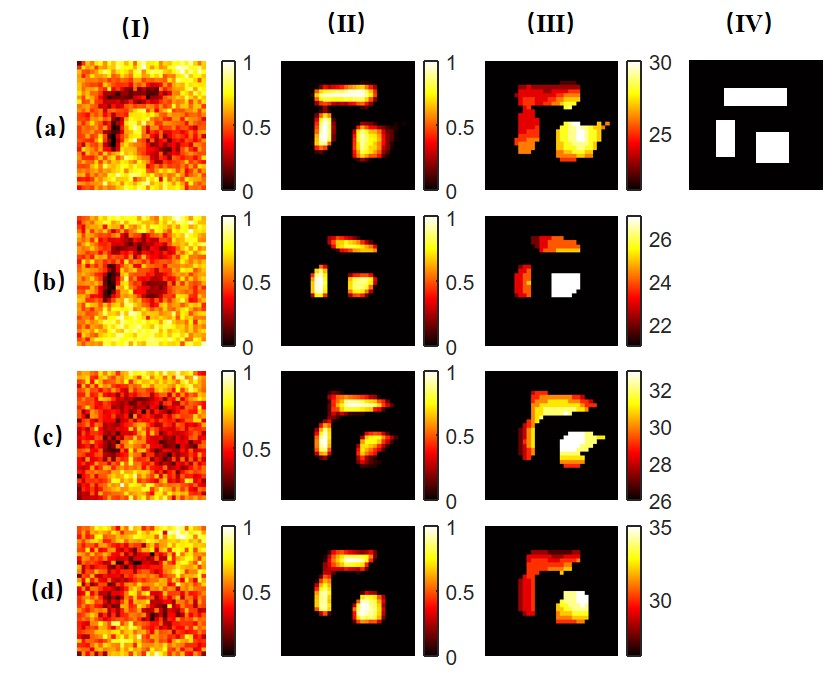}
\caption{Experimental results with negative targets hidden in three depths. Row (a) to (d) shows the reconstructed maps for different depths, respectively. The The hidden depth is ranging from 5 mm to 30 mm for each target. Column (I) to (III) gives the single pixel recovery results, intensity map, and depth map, respectively. Column (IV) displays the ground truth. }
\label{fig:negative_three_depth}
\vspace{-10pt}
\end{figure}

To further showcase the capacity of continuous depth imaging and simultaneous imaging of the deep, middle, and shallow layers offered by the proposed method, a three-layer negative target consisting of simple geometric objects is constructed. Specifically, a short strip, a long strip, and a square are located at three distinct depths respectively. The short strip is located in the shallow layer. The long strip is buried in the middle layer. And the square is buried in the deep layer. The depths for each object in each group is listed in Table. \ref{tab:my_table}. Fig. \ref{fig:negative_three_depth} shows the reconstruction result. Fig. \ref{fig:negative_three_depth} row (a) to (d) illustrates the reconstructed maps for each scenario. The single pixel recovery results are given in Fig. \ref{fig:negative_three_depth} column (I). The depth maps and intensity maps are shown in Fig. \ref{fig:negative_three_depth} column (II) and column (III), respectively. As we can see from Fig. \ref{fig:negative_three_depth}, the intensity maps and depth maps are successfully reconstructed. More importantly, the imaging performance is satisfactory.

\section{Discussion}
Most of current methods for 3D imaging into/through scattering medium can be classified into a scheme combining focal plane 2D imaging and depth discrimination. Optical memory effect based methods like speckle correlation imaging are typical ones that acquire an 2D image in a specific plane via array detectors, such as charge-coupled device (CCD). Then the depth discrimination is realized by the calibrated geometry of the optical system. Calibration is often invasive, complex and limited scope of application. It should be pointed out that optical memory based methods are not applicable to thick scattering medium. All the experiments conducted in memory effect based methods are omitting a truth that the affect of scattering on objects locates different depth differs. While in memory effect based methods, the scattering is induced by only one diffuser. Thus the photon scattering suffered is the same for the objects locating in different depths. Our proposed SP$^3$ 3DI needs no calibration, prior and reference. If higher repeated frequency laser is considered, for example, 80 MHz, the measurement acquisition process will be much faster compared with our 1 MHz repeated frequency laser.

Other direct method for 3D imaging into/through scattering medium is difficult to solve transport equation. Time of light is applicable for diffusive tomography which depicts the propagation of light wave. However, solving diffusive equation is computation expensive and time consuming.  

Binocular vision is a method intensively applied in 3D imaging in free space. It needs images with high quality. However, in scattering medium, the reconstructed image is not that good or even poor. Current 3D imaging methods based on binocular vision for imaging through/into scattering medium are laying the foundation on the good reconstruction quality which is realized by optical memory effect. Thus, it is not appropriate for imaging through/into thick scattering medium. 

\section{Conclusion}
In this manuscript, we present a simple yet powerful 3D imaging scheme, SP$^3$ 3DI, for imaging through unknown thick scattering medium. Both tradition method and deep learning-based method are proposed. The two proposed methods demonstrate a satisfying capability for 3D imaging. Additionally, the multi-depth discrimination of SP$^3$ 3DI highlights its impressive depth of focus. The low computational complexity of the deconvolution process allows for rapid imaging with SP$^3$ 3DI fast imaging. SP$^3$ 3DI realizes 3D imaging through unknown thick scattering medium with only one single pixel detector. Furthermore, SP$^3$ 3DI indicates the capability of simultaneously imaging shallow and deep, which is rare for imaging into/through scattering medium.

\section*{Funding}
National Natural Science Foundation of China (62305302); Zhejiang Lab Startup Fund (113010-PI2107).

\section*{Disclosures}
The authors declare no conflicts of interest. 

\section*{Data availability}
Data underlying the results presented in this paper are not publicly available at this time but may be obtained from the authors upon reasonable request.


\bibliography{sample}

\begin{table}[p]
\centering
\caption{\bf Hidden depths of the three-layer targets}
\label{tab:my_table}
\begin{tabular}{c|c|c|c|c} \hline
Short Strip& 5 mm& 5 mm& 15 mm& 15 mm
\\
Long strip& 10 mm& 15 mm& 20 mm& 25 mm
\\ 
Square& 20 mm& 20 mm& 30 mm& 30 mm
\\ \hline
\end{tabular}
\end{table}

\end{document}